\documentclass[twocolumn, superscriptaddress, prl, amsmath, amssymb, aps, 10pt, floatfix]{revtex4-2}

\usepackage{float}
\usepackage{graphicx}
\usepackage{dcolumn}
\usepackage{bm}
\usepackage{xcolor}
\usepackage{textgreek}
\usepackage{textcomp}
\usepackage[separate-uncertainty=true, multi-part-units=single, list-units=repeat]{siunitx} 
\usepackage[version=4]{mhchem} 
\usepackage[thinc]{esdiff} 
\usepackage{comment}
\usepackage{textcomp}
\usepackage{microtype}

\begin{document}

\preprint{APS/123-QED}

\title{Discharge at the Microscale: Using Optical Tweezers to Observe Muon-Induced Discharges of a Levitated Microparticle in Air}

\author{Andrea Stoellner}
\affiliation{Institute of Science and Technology Austria, Am Campus 1, 3400 Klosterneuburg, Austria}
\author{Isaac C.D.~Lenton}
\affiliation{Institute of Science and Technology Austria, Am Campus 1, 3400 Klosterneuburg, Austria}
\author{Caroline Muller}
\affiliation{Institute of Science and Technology Austria, Am Campus 1, 3400 Klosterneuburg, Austria}
\author{Scott Waitukaitis}
\email{scott.waitukaitis@ista.ac.at}
\affiliation{Institute of Science and Technology Austria, Am Campus 1, 3400 Klosterneuburg, Austria}

\date{\today}

\begin{abstract}

Electrical discharge at the smallest possible length and charge scales is not well understood. Using optical tweezers, we investigate spontaneous discharges of a single micron-scale particle levitated in air. These ``microdischarges'' have a typical size of $\sim$40 $|e|$, but can be as small as a few $|e|$ and as large as several hundred. The absence of a well-defined trigger charge and the weak dependence on particle size suggest events are not classical gaseous breakdown. 
Instead, we show that microdischarge events arise from the rapid capture of ions left in the tracks of nearby passing ionizing radiation. Our results highlight the role of natural ionizing radiation in initiating micron-scale discharges and provide a platform for studying discharge physics in electrode-free environments and at the smallest scales.
\end{abstract}

\maketitle

The physics of how objects lose electrical charge is important in many contexts. In gases at the macroscale, discharge is most commonly considered in the context of dielectric breakdown mediated by Townsend avalanches~\cite{Paschen1889, Townsend1900}. Breakdown experiments typically use pairs of metal electrodes and probe the underlying physics through the current that flows between them. When electrode separations shrink to tens of microns or less, gaseous breakdown becomes more difficult to initiate, and alternative discharge mechanisms become important, \textit{e.g.}, field emission \cite{Boyle1955, Hourdakis2006, Peschot2014, Loveless2017, Fu2020, Wang2022}. The focus on electrode pairs and current measurements has left a different regime of discharge largely unexplored -- namely, what is expected to occur with an isolated, highly charged micro- or nanoparticle. While such objects can be expected to experience discharge, potential mechanisms governing this remain unclear, and need not correspond to what occurs at larger scales. Such a regime would be relevant, for example, to the highly charged microparticles and droplets that comprise a thundercloud, where it has been suggested to play a role in the initiation of lightning \cite{Gurevich2013, Dwyer2014, Petersen2015, Babich2017}. However, direct observation has remained out of reach, as existing approaches have not been able to probe such small length and charge scales.

In this work, we use optical tweezers to simultaneously charge and measure the charge of an isolated microparticle levitated in air, allowing us to observe individual “microdischarges” in which the particle suddenly loses charge. By monitoring individual particles over weeks at a time, we gather comprehensive statistics that resolve the distributions of microdischarge sizes (from a few to several hundred $|e|$), the trigger charge values at which events occur, and the time between events. The absence of a well-defined threshold in trigger charge and weak dependence on particle size suggest that the observed events do not arise from classical dielectric breakdown. Instead, through simultaneous detection of cosmic-ray muons traversing our chamber, we find that microdischarges occur via the rapid capture of ions produced along tracks of ionizing radiation. Our results introduce a new experimental regime of discharge physics involving isolated micro- and nanoscale particles, countable numbers of elementary charges, and charge loss driven by the capture of ions generated along ionization tracks from nearby passing radiation.

\begin{figure}[htpb!]
\centering
\includegraphics{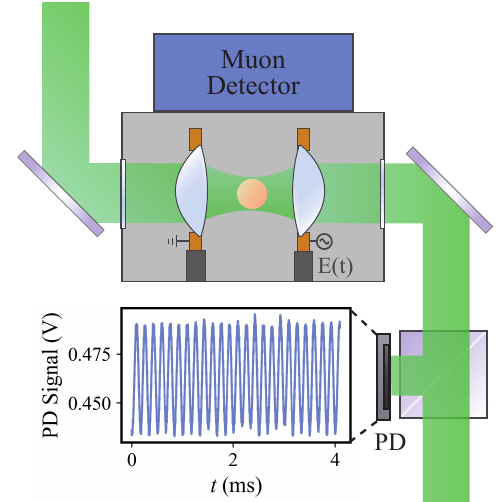}
\caption{ \label{fig:experimental_setup} Experimental setup. An SiO\textsubscript{2} sphere with a diameter of $\sim$1~\textmu{m} is held by a counter-propagating optical trap ($\lambda~=~$~532~nm) inside a grounded metal chamber. Two copper ring electrodes around the trapping lenses generate an AC electric field, causing the charged particle to oscillate along the beam axis. A beamsplitter on one side of the chamber directs the light to a photodiode (PD), which records the scattered light coming from the particle carrying information about the particle's position. The inset on the bottom of the figure shows a typical oscillating signal recorded by the PD. On top of the chamber a muon detector equipped with two 5 cm x 5 cm scintillators continuously records incoming muon events.}
\end{figure}

The experimental setup is illustrated in Fig.~\ref{fig:experimental_setup} (for further details see \cite{Stoellner2025}). A 532~nm continuous-wave laser forms a counter-propagating optical trap with 40~mW power inside a grounded metal chamber. The trapped particle is an amorphous SiO\textsubscript{2} sphere with a diameter of 1.18~\textmu{m} for the experiments presented in the main text. The two trapping lenses (8~mm focal lengths) are enclosed by copper ring electrodes with $\sim$9~mm separation. By applying an AC voltage (500~V at a frequency $f_{dr} =$ 6~kHz) to one of the ring electrodes while keeping the other one grounded, an alternating electric field of $\sim$45~kV/m ($\textless$~2~\% of the dielectric strength of air) is applied to the trap. If the trapped particle carries electric charge, it oscillates along the beam axis in response to the electric field, as shown in the inset of Fig.~\ref{fig:experimental_setup}. This motion is recorded by a photodiode (sampling frequency 1~MHz) in the beam path on the right-hand side of the chamber. On top of the chamber there is a muon detector (UKRAA PicoMuon) equipped with two 5~cm~x~5~cm scintillators stacked on top of each other. While the detector is sensitive to any kind of ionizing radiation that passes through the scintillators, a muon event is only recorded if both scintillators are triggered simultaneously.

\begin{figure}[htpb!]
\centering
\includegraphics{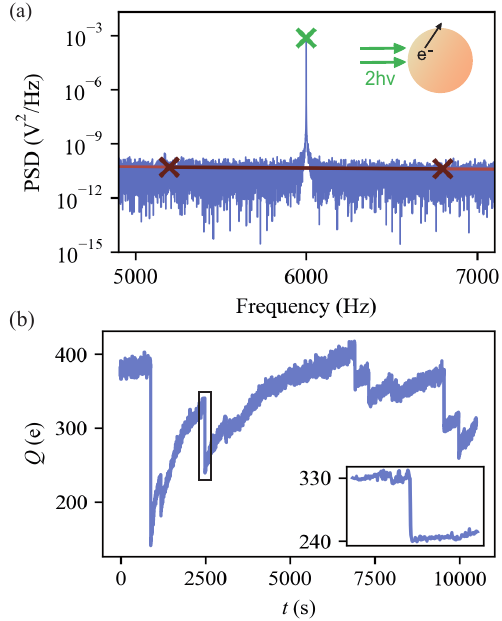}
\caption{ \label{fig:discharges} Measurement principle and charge curve showing spontaneous discharges. (a) The particle charge is calculated from the ratio of the height of the charge peak at the driving frequency of 6 kHz (green cross) to the thermal background of the power spectral density (PSD). The thermal background value is calculated from the geometric mean of two measurement points $\pm$ 800~Hz from the charge peak (brown crosses). The particle continually gains charge by a two-photon process induced by the trapping laser \cite{Stoellner2025}. When two photons combine, they carry enough energy to emit an electron from the particle, therefore giving it +1~$|e|$. (b) We record the particle charge continuously over several days to weeks without manipulating the system, frequently observing spontaneous discharges. The inset highlights one particular event of $\sim$90~$|e|$. 
}
\end{figure}

To measure the trapped particle's charge, we calculate the power spectral density (PSD) of its motion from the photodiode signal, as shown in Fig.~\ref{fig:discharges}(a). The particle charge can be extracted from the ratio between the charge peak $S_{el}$ and the thermal background $S_{th}$ at the driving frequency $f_{dr}$ \cite{Ricci2019}:

\begin{equation}
    Q \propto \sqrt{\frac{S_{el}(f_{dr})} {S_{th}(f_{dr})}},
    \label{eq:charge equation}
\end{equation}
In a lock-in style measurement we record $S_{el}$ directly while estimating $S_{th}$ from the geometric mean of the two PSD values $\pm$ 800 Hz from the charge peak. We continuously record the thus calculated particle charge at a rate of five measurements per second. As we elaborate in a prior publication, the trapping laser itself causes the particle to charge via two-photon electron ejection \cite{Stoellner2025}.  This mechanism results in a steady increase of positive charge on the particle surface, allowing us to reliably and continually charge the particle to high levels. Fig.~\ref{fig:discharges}(b) shows a typical charge curve with steady charging by the laser, as well as sudden drops, \textit{i.e.}, microdischarges. This behavior occurs consistently over several weeks in a closed-off chamber without external intervention. The microdischarges can range from just a few elementary charges up to \textgreater200~$|e|$ and occur within a single measurement window, \textit{i.e.}, within 0.2~s, although larger events occasionally appear to span two to three measurement windows. 

\begin{figure}[htpb!]
\centering
\includegraphics{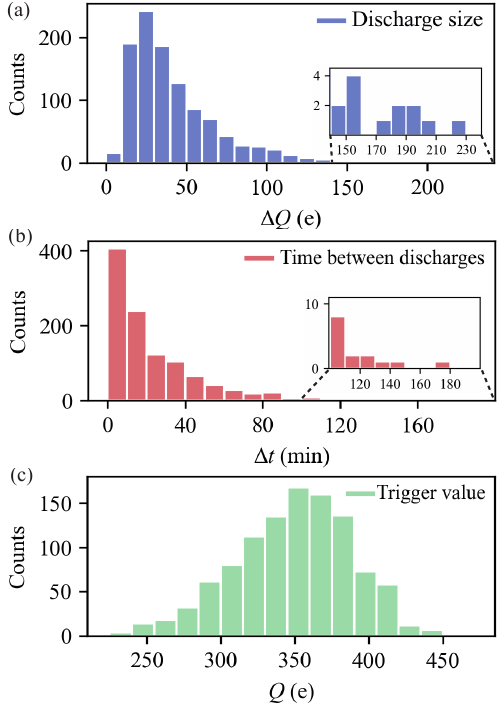}
\caption{ \label{fig:statistics} Microdischarge statistics of a particle with a diameter of $~1.18$~\textmu{m} undergoing 1076 events over a period of two weeks. (a) Distribution of the difference in particle charge before and after individual microdischarges, showing charge drops of up to $\sim$230~$|e|$. (b) Distribution of the time between individual events in minutes. As we show in the Supplemental Material \cite{SupplMat}, this distribution is exponential and therefore indicative of a Poisson process. (c) Distribution of the trigger values at which each of the microdischarges occurred.}
\end{figure}

Such microdischarges occur whenever we catch a particle and allow it to become highly charged in our optical trap.  For this paper, we focus on results from one trapped particle, which we continuously monitored over a two week period, during which we recorded 1076 events. Statistics corresponding to these events are shown in Fig.~\ref{fig:statistics}. Panel~(a) shows the distribution of microdischarge sizes, defined as the difference between the mean particle charge in the three seconds before and three seconds after the event. Most microdischarges were between 10~$|e|$ and 40~$|e|$, although significantly larger events were also observed. The smallest detected event was 6 $|e|$ while the largest was 224~$|e|$ -- nearly two thirds of the mean particle charge of 352~$|e|$. As shown in Fig.~\ref{fig:statistics}(b) and discussed in more detail in the Supplemental Material~\cite{SupplMat}, the time intervals between individual events follow an exponential distribution. This behavior is characteristic of a Poisson process, indicating that the events occur randomly in time with a constant average rate (one every $\sim$25 minutes) and are statistically independent of one another. Finally, panel~(c) shows the distribution of charge values at which an event was triggered (\textit{e.g.}, 330~$|e|$ for the event shown in the inset of Fig.~\ref{fig:discharges}(b)), with the mean trigger value of 348~$|e|$ being very close to the mean particle charge over the measurement period. The lowest and highest recorded trigger values were 168~$|e|$ and 444~$|e|$, respectively. 

It is natural to wonder if the microdischarges we observe arise from electrical breakdown of the air surrounding the particle, however several observations are inconsistent with this interpretation. The mean trigger charge corresponds to a surface electric field of $1.4~\mathrm{MV/m}$. While this is large, it falls short of the nominal dielectric strength of air, $E_b \approx 3~\mathrm{MV/m}$ in standard conditions. Moreover, it is well established that the dielectric strength increases when breakdown occurs over length scales of $\sim$10~\textmu m and smaller \cite{Boyle1955, Hourdakis2006, Fu2020}, and in our situation the region of high electric field is restricted to a few hundred nanometers. A second discrepancy arises from the distribution of trigger charges, which does not exhibit a clear lower threshold but instead mirrors the overall distribution of particle charge values (see Supplemental Material \cite{SupplMat}). Finally, microdischarge data from particles of different diameters (0.69–1.86~\textmu m; see Supplemental Material \cite{SupplMat}) show no prominent size dependence, despite substantial variation in surface electric field. 

An alternative explanation for the observed microdischarges is that they correspond to events in which ionizing radiation passes near to the particle. Possible sources include those from radioactive elements (\textit{e.g.}, radon) \cite{Cooper1973, Adachi1983, Bazilevskaya2008, Mishev2022} or cosmic radiation. Upon initial consideration, radon decay might seem the leading candidate, since it is a significant contributor to air ionization at sea level \cite{Wilkening1988, Zhang2011}. As we discuss in the Supplemental Material, however, typical indoor levels of radon decay correspond to a frequency of events that is too small to correspond to our observations.

\begin{figure}[htpb!]
\centering
\includegraphics{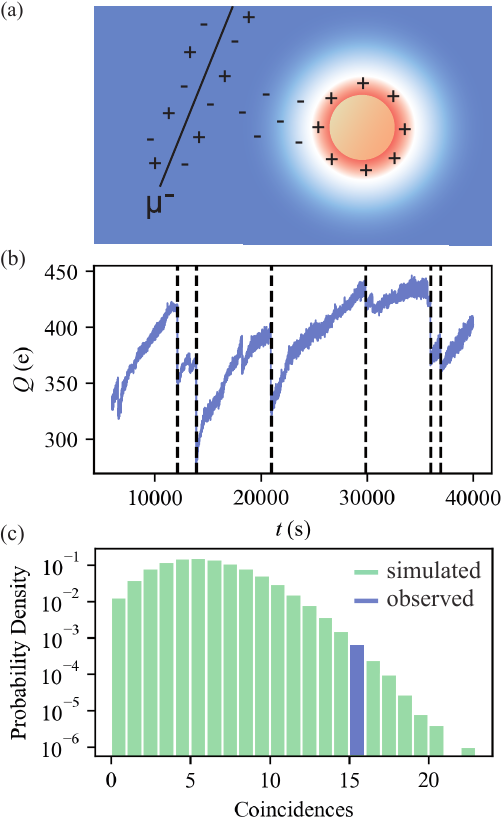}
\caption{ \label{fig:muons} Link between cosmic-ray muons and microdischarges. (a) An illustration of the proposed mechanism showing the ionization track left behind by a muon close to the particle. Negative ions are then drawn in by the positively charged particle's electric field, leading to a sudden loss of charge. (b)~Charge \textit{vs.}~time curve showing microdischarges that coincided with a muon event (dashed lines). Not all events are associated with a measured muon passing. (c) Probability density of the number of microdischarges that coincide with a muon event for $10^6$ randomly generated trials (green distribution) and 58 experimentally observed microdischarges (blue bin). For the random trials we generated artificial microdischarge times, following the Poisson statistics of the experimental data. For each trial run the number of discharges that coincide with a muon event is recorded, giving an estimate of how likely it is to observe a muon-discharge event coincidentally. Only $\sim$$10^3$ of the $10^6$ randomly generated trials yielded $\ge$ 16 coincidences, resulting in a p-value of $p = 0.001$.
}
\end{figure}

We therefore work with the hypothesis that the observed microdischarges are caused by cosmic radiation, and in particular by cosmic-ray muons. As the dominant component of cosmic radiation at the Earth’s surface~\cite{Gaisser2016, Autran2018}, and owing to their ability to easily penetrate solid materials (\textit{e.g.}, our chamber walls), high-energy muons are a natural candidate to consider. A sketch of how muons might discharge the particle is shown in Fig.~\ref{fig:muons}(a). As a muon passes through the air, it leaves behind a track of positive and negative ions \cite{Bazilevskaya2008, Mishev2022}. If this track is sufficiently close to the particle, negative ions can be drawn in by the particle’s electric field, leading to a sudden drop in charge. To test this hypothesis, we compare simultaneous time series of particle charge and muon events detected by the muon detector mounted on top of our chamber over the course of several days. Using these data, we determine the number of discharges that occur in close temporal proximity ($\pm$0.2~seconds, \textit{i.e.}, our charge-measurement window) to a muon event, hereafter referred to as a “coincidence”. Fig.~\ref{fig:muons}(b) shows a segment of a charge \textit{vs.}~time curve with dashed lines marking coincidences where muon events were recorded together with a discharge. There are many more muon events than discharges ($\sim$1 every 6 seconds \textit{vs.} 1 every 25 minutes), which is expected since most muons passing through the relatively large area of our detector (25 cm$^2$) will not pass near the particle. There are also more discharges than coincidences: roughly one in four discharge events has an associated muon detection. As will be explained shortly, both of these factors can be accounted for by our muon hypothesis.

To establish a statistically significant case for causality between discharge and muon coincidences, we perform numerical simulations to determine how likely our observations are to arise by chance. In these simulations, we consider a fixed observation window (72 hours), during which we recorded $\sim$53,000 muon events, 58 discharges, and 16 coincidences. We then ask: if the timestamps of discharges were randomly redistributed within this time window, what is the probability of obtaining different numbers of coincidences? To estimate this probability, we perform $10^6$ stochastic trials in which a Poisson-distributed number of discharge events with a mean count of $\mu=$~58~discharges is assigned within the measurement interval, while the experimental muon data remain unchanged. As before, coincidences between (synthetic) discharges and muon events are counted when their temporal difference is less than 0.2 seconds. The results are shown in Fig.~\ref{fig:muons}(c), where the $y$ axis is plotted on a logarithmic scale. On average, only about six out of $\mu=$ 58 discharges coincide with a muon event if there is no causal relationship between the two. The probability of observing 16 or more coincidences out of 58 discharges corresponds to only $\sim$$10^3$ out of $10^6$ randomly generated trials, resulting in a p-value of 0.001. This result highlights the strong correlation in our experimental datasets: the observed “coincidences” correspond to causal events between muon detection and subsequent discharge. We reach a similar conclusion when the simulations are repeated using discharge timestamps drawn from other time windows of the experiment (rather than synthetically generated times), ruling out the possibility that a hidden temporal pattern in the discharge sequence produces the observed correlation. Moreover, several other factors support this picture.  First, the muon events, like the discharge events, exhibit Poisson statistics (see Supplemental Material \cite{SupplMat}). Second, considering our muon detector only covers about 1/4th of the total solid angle from which muons can come (see Supplemental Material \cite{SupplMat}, we expect only about 25 $\%$ of discharge events to have an associate muon event -- which is very close to what we observe (27 $\%$).

In summary, we have used optical tweezers to study a new regime of discharge physics corresponding to a single, highly charged microparticle levitated in air. By continuously charging and monitoring the particle’s charge, we observe discrete microdischarge events consisting of a few to a few hundred elementary charges. The absence of a well-defined threshold in trigger charge, the fact that the trigger distribution simply mirrors the underlying charge distribution, and the weak dependence on particle size suggest that the observed events do not arise from classical field-driven gaseous breakdown. Instead, we find that they correspond to the sudden attraction of ions created in the tracks of nearby traversing cosmic-ray muons. We reach this conclusion by examining the temporal coincidence between muon and discharge events and showing that the observed number of coincidences is highly unlikely to arise by chance alone. In this sense, the observed microdischarges are more closely related to the ionization “seeding” events that initiate classical breakdown, as opposed to the Townsend Avalanche mechanism itself. Our findings bring to light a distinct regime of discharge physics at the smallest length and charge scales, in which ionization from natural radiation can suddenly and dramatically influence particle charge. This regime has direct relevance to fields such as aerosol and atmospheric physics, where the dynamics of charged particles play crucial roles.

\vspace{1em}
We thank Todor Asenov and Abdulhamid Baghdadi for their outstanding technical support. This project has received funding from the European Research Council (ERC) under the European Union’s Horizon 2020 research and innovation programme (Grant agreements No.~949120 and No.~805041). This research was supported by the Scientific Service Units of The Institute of Science and Technology Austria (ISTA) through resources provided by the Miba Machine Shop.

\bibliographystyle{apsrev4-2}
\bibliography{manuscript}

%

\end{document}


\setcounter{figure}{0}
\renewcommand{\figurename}{SUPPL.~FIG.}
\renewcommand\thefigure{\arabic{figure}}
\setcounter{table}{0}
\renewcommand{\tablename}{SUPPL.~TABLE}
\renewcommand\thetable{\arabic{table}}
\renewcommand{\theequation}{S\arabic{equation}}
\setcounter{equation}{0}  

\preprint{APS/123-QED}

\title{Supplemental Information \\
Discharge on the Microscale: Using Optical Tweezers to Observe Muon Induced Discharges of a Levitated Microparticle in Air}

\author{Andrea Stoellner}
\affiliation{Institute of Science and Technology Austria, Am Campus 1, 3400 Klosterneuburg, Austria}
\author{Isaac C.D. Lenton}
\affiliation{Institute of Science and Technology Austria, Am Campus 1, 3400 Klosterneuburg, Austria}
\author{Caroline Muller}
\affiliation{Institute of Science and Technology Austria, Am Campus 1, 3400 Klosterneuburg, Austria}
\author{Scott Waitukaitis}
\email{scott.waitukaitis@ista.ac.at}
\affiliation{Institute of Science and Technology Austria, Am Campus 1, 3400 Klosterneuburg, Austria}

\date{\today}


\maketitle

\section{Trigger value vs particle charge}

\begin{figure}[H]
\centering
\includegraphics{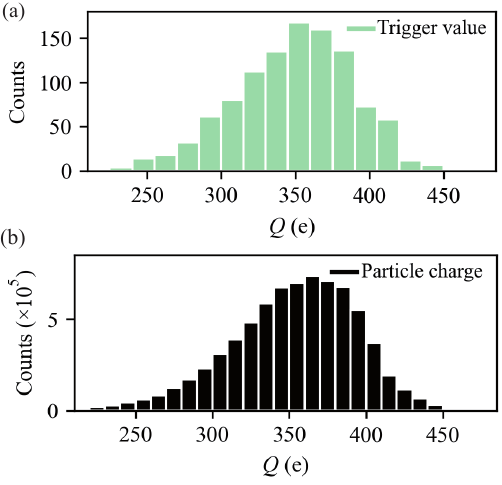}
\caption{\label{fig:SI particle charge} Comparison between (a) values at which discharges were triggered and (b) all particle charge values over the course of the measurement. The strong resemblance between the two distributions indicates that discharges are not tied to a particular charge value, but instead reflect the charge states in which the particle spends most of its time.
}
\end{figure}

The charge values at which discharges are triggered (trigger values) do not denote a threshold at which discharge becomes possible, but instead reflect the particle charge over the measurement period. This can be clearly seen from Suppl.~Fig.~\ref{fig:SI particle charge} which compares the distribution of (a) discharge trigger values with (b) all measured charge values over the course of the experiment. The distribution of charge values itself arises from a balance between the continuous positive charging by the laser and the discharges, which in turn reduce the particle charge. The skew of the distribution towards lower values (some outliers as low as 168~$|e|$ for the trigger values and 95~$|e|$ for the charge values) can be explained by discharges that happen closely after each other, as well as the nature of the two-photon laser charging. As we see in Fig. 3(b) in the main manuscript, it is much more likely to have discharges shortly one after another than to have a long interval without discharges. For discharges with a small $\Delta t$, the particle has little time to recover the lost charge and will therefore undergo the second discharge at a lower than usual trigger value. Conversely, in the already much more infrequent case of a particularly long $\Delta t$, the particle has more time to acquire charge and, as a result, the expected trigger value would indeed be higher. However, as the laser charging becomes progressively slower as the particle charge increases, very high trigger values become even rarer, explaining the distribution's skew towards lower values. 

\section{Different particle sizes}

In order to test a potential dependency of the discharges on the particle's surface electric field, we have measured discharges of particles ranging in size from 0.69~\textmu{m} to 1.86~\textmu{m}. Their corresponding mean surface electric fields ranged from $E = 1.3$~$\mathrm{MV/m}$ for the biggest size up to $E = 4.3$~$\mathrm{MV/m}$ (maximum recorded: $E = 5.2$~$\mathrm{MV/m}$) for the smallest size. We do not observe a significant difference in either the size or the frequency of the discharges, as can be seen from Suppl. Fig. \ref{fig:SI particle sizes}(a) and (b). The trigger values shown in panel (c) do not increase with the particle radius $r^2$, as one would expect if the discharges were triggered by the particle's electric field, but instead reflect the mean charge the particle had during the measurement. 

\begin{figure}[H]
\centering
\includegraphics{fig_SI_particle_discharge_size.pdf}
\caption{\label{fig:SI particle sizes} Comparison of (a) discharge sizes, (b) inter-discharge times, and (c) trigger charge values for different particle sizes ranging from 0.69~\textmu{m} to 1.86~\textmu{m}.
}
\end{figure}

\section{Estimation of cross section}

\begin{figure}[H]
\centering
\includegraphics{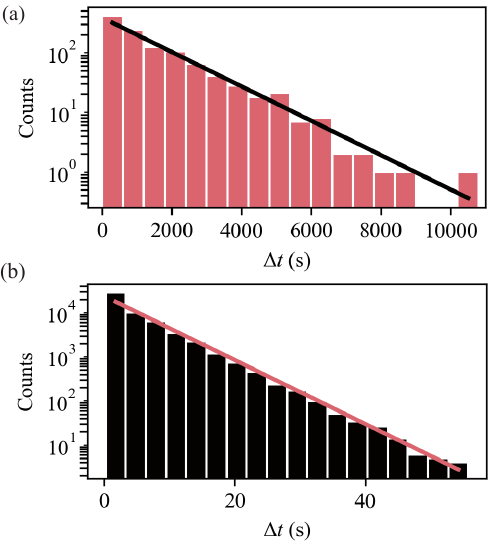}
\caption{\label{fig:SI exponential} Exponential nature of (a) the time between discharge, and (b) the time between muon events.
}
\end{figure}

As our discharge mechanism requires cosmic-ray muons to pass close to our levitated particle, we can infer an estimate of how close they need to come to trigger a discharge from a comparison between our measured muon rate and discharge rate. Suppl.~Fig.~\ref{fig:SI exponential} shows the distribution of inter-event times for (a) discharges and (b) muons, exhibiting the exponential decay expected for a Poisson process. From the respective fits we find the time constants $\tau_{discharge} = 1509$~s and $\tau_{muon} = 6$~s. As the incident muon rate $\lambda = 1/\tau$ is proportional to the cross section $\sigma$, we find an inverse proportional relationship between the two time constants and the two corresponding cross sections:

\begin{equation}
    \frac{\tau_{1}}{\tau_{2}} = \frac{\sigma_{2}}{\sigma_{1}}.
    \label{eq:cross section}
\end{equation}
Knowing the area of the muon detector's scintillators of 25~$\mathrm{cm^2}$ and the two time constants, we find the cross section which a muon must pass in order to cause a discharge is $\sim$10~$\mathrm{mm^2}$, corresponding to a disk with a radius $r=1.8~\mathrm{mm}$. This value likely reflects an interplay between the particle's electric field, which is always attractive for negative ions and the much lower, yet larger-scale AC electric field we use to measure the particle charge, which displaces ions by up to $\sim$1.8~mm in one phase ($E_{AC} \sim45~\mathrm{kV/m}, f_{AC} = 6~\mathrm{kHz}$, average negative ion mobility $\mu = \mathrm{2.4~cm^2~V^{-1}~s^{-1}}$ \cite{Nagato1988}). A similar enlargement of the cross section is conceivable for other background electric fields, \textit{e.g.}, the electric field of a thundercloud.

We can do the same estimation for ions released by radon decay with an average indoor level of radon decay in Austria of $\mathrm{112~Bq/m^3}$ \cite{radonEurope}. To achieve our measured discharge rate, the particle would need to capture ions from decays in a volume of $\mathrm{5.9~cm^3}$, corresponding to a sphere with radius $r=1.1~\mathrm{cm}$. As this region extends beyond the volume enclosed by the electrodes used to measure the particle charge, ions produced by radon decay would be expected to drift toward the electrodes rather than reach the particle. Radon decay therefore cannot explain our observed discharge frequency.

\section{Estimation of coincidence rate}

As discussed in the main manuscript, we can attribute about one in four discharge events to a muon detection. To understand this value, we consider the solid angle of our muon detector given by
\begin{equation}
    \Omega \approx \frac{A}{d^2},
    \label{eq:solid angle}
\end{equation}
where $A=\mathrm{25~cm^2}$ is the detector area and $d\sim7$~cm is the distance between the detector and our trapped particle, resulting in a solid angle of $\Omega=0.5~\mathrm{sr}$. In addition, we have to consider the angular distribution of incoming muons at sea level, which is generally approximated by 
\begin{equation}
    I(\theta) = I_0 cos^2(\theta),
    \label{eq:muon distribution}
\end{equation}
where $\theta$ is the zenith angle and $I_0$ is the flux at $\phi=0$, i.e. from directly above. By integrating the angular distribution over the solid angle covered by the detector we find that the expected muon flux through the detector is $\sim$25~\% of the overall flux, which is in agreement with our observations.

\bibliographystyle{apsrev4-2}
\bibliography{supplement}